\documentclass[twocolumn,english,prl,reprint,superscriptaddress]{revtex4-1}
\usepackage[T1]{fontenc}
\usepackage[latin9]{inputenc}
\usepackage{amsmath}
\usepackage{amssymb}
\usepackage{graphicx}

\makeatletter
\@ifundefined{textcolor}{}
{%
 \definecolor{BLACK}{gray}{0}
 \definecolor{WHITE}{gray}{1}
 \definecolor{RED}{rgb}{1,0,0}
 \definecolor{GREEN}{rgb}{0,1,0}
 \definecolor{BLUE}{rgb}{0,0,1}
 \definecolor{CYAN}{cmyk}{1,0,0,0}
 \definecolor{MAGENTA}{cmyk}{0,1,0,0}
 \definecolor{YELLOW}{cmyk}{0,0,1,0}
}

\usepackage[english]{babel}
\def\urlprefix{}
\def\url#1{}
\addto\captionsenglish{%
}

\makeatother

\usepackage{babel}
\begin{document}

\title{Topological Spin Hall Effect due to Magnetic Skyrmions}

\author{Gen Yin}

\thanks{E-mail: gyin001@ucr.edu}

\affiliation{Department of Electrical Engineering, University of California, Riverside,
CA 92521-0204, USA}

\author{Yizhou Liu}

\affiliation{Department of Electrical Engineering, University of California, Riverside,
CA 92521-0204, USA}

\author{Yafis Barlas}

\affiliation{Department of Electrical Engineering, University of California, Riverside,
CA 92521-0204, USA}
\affiliation{Department of Physics and Astronomy, University of California, Riverside,
CA 92521-0204, USA}

\author{Jiadong Zang}

\thanks{corresponding author; E-mail: jiadongzang@gmail.com}

\affiliation{Department of Physics and Astronomy, Johns Hopkins University, Baltimore,
MD 21218, USA}

\author{Roger K. Lake}

\thanks{corresponding author; E-mail: rlake@ee.ucr.edu}

\affiliation{Department of Electrical Engineering, University of California, Riverside,
CA 92521-0204, USA}
\begin{abstract}
The intrinsic spin Hall effect (SHE) originates from the topology of the Bloch bands in momentum space.
The duality between real space and momentum space calls for a spin Hall effect induced from a real space topology
in analogy to the topological Hall effect (THE) of skyrmions.
We theoretically demonstrate the topological spin Hall effect (TSHE) in which a pure transverse spin
current is generated from a skyrmion spin texture.
\end{abstract}
\maketitle
Transverse spin accumulation in semiconductors due to extrinsic spin-orbit
scattering was first predicted by Dyakonov and Perel \cite{dyakonov_current-induced_1971,dyakonov_possibility_1971}.
Strong spin-orbit coupling (SOC) of the disorder scatters different
spins in opposite directions leading to a non-zero transverse spin
current perpendicular to the charged current. 
Evidences of the predicted asymmetric scattering of different spins was later abserved in optical \cite{bakun_observation_1984} and photovoltaic \cite{tkachuk_resonant_1986} 
experiments. 
Hirsch named this phenomenon the `spin Hall effect' (SHE) and
proposed that the chargeless transverse spin current can be
transferred back to a Hall voltage using an inverse SHE measurement
\cite{hirsch_spin_1999}.
Later theoretical
studies predicted an intrinsic contribution to the SHE  in the presence of SOC due to the topological property of the Bloch states at the
Fermi surface \cite{murakami_dissipationless_2003,sinova_universal_2004,nikolic_nonequilibrium_2005,grover_topological_2008,iwasaki_universal_2013}.
Direct observations of the SHE have been experimentally achieved in semiconductors using
Kerr rotation microscopy\cite{wunderlich_experimental_2005,kato_observation_2004}.

In magnetic materials due to SOC, extrinsic or intrinsic mechanisms can lead to a
non-linear contribution to classical Hall signal\cite{smit_spontaneous_1955,berger_side-jump_1970,onoda_topological_2002}.
The non-linearity which is proportional to the magnetization is a result of the
transverse accumulation of itinerant majority spins resulting in the anomalous Hall effect (AHE)
\cite{nagaosa_anomalous_2010}. Similar to the SHE, the AHE can result from an intrinsic or
extrinsic mechanism. The intrinsic contribution to the AHE is related to the
Berry curvature within the Fermi surface, which is determined by the
topological nature of the Bloch bands\cite{haldane_berry_2004,onoda_topological_2002}.

The momentum-space topological origin of the intrinsic AHE is
the same to that of the intrinsic SHE. Similarly, the real-space topology
of a magnetic system can also induce a Hall effect \cite{volovik_fractional_1989}. An electron hopping
through magnetic sites with particular chiral textures acquires a
Berry phase and thus experiences an emergent gauge field during transport
\cite{taguchi_spin_2001}.
The emergent gauge field generates a Hall voltage that does not
originate from SOC, which is usually referred to as the `topological
Hall effect' (THE) \cite{bruno_topological_2004}.
Recently, a skyrmion lattice, a topologically non-trivial chiral
spin texture, has been observed in helical magnets with
a Dzyaloshinskii-Moriya (DM) interaction
\cite{yu_real-space_2010,muhlbauer_skyrmion_2009,schulz_emergent_2012}.
These materials provide robust samples where the THE has been
detected, and the measured Hall signal is a signature of the
skyrmion phase in many B20 magnetic compounds
\cite{li_robust_2013,kanazawa_large_2011,neubauer_topological_2009,huang_extended_2012}.

%
In the adiabatic limit, each electron spin passing through a single
skyrmion has its spin aligned with the direction of  spatial
magnetization of the skyrmion which generates an emergent gauge
field of up to one flux quantum \cite{schulz_emergent_2012}.
This flux quantum confined in the area of a single skyrmion gives a
gigantic effective field, that makes the THE a possible detection
method for skyrmions.
Moreover, the direction of the local magnetic field generated by
this emergent gauge field is opposite for parallel and antiparallel
spin, which deflects them in opposite directions.
This might separate the spin current from the charge current,
generating an unconventional topological spin Hall effect (TSHE)
which does not originate from band topology.
Motivated by these
possibilities, in this letter we theoretically investigate the THE
and the TSHE resulting from a single magnetic skyrmion.
The  TSHE phenomenon discovered here can be explained in terms of a
general physical picture that would apply equally
well to a skyrmion lattice.

\begin{figure}
\begin{centering}
\includegraphics[width=0.8\columnwidth]{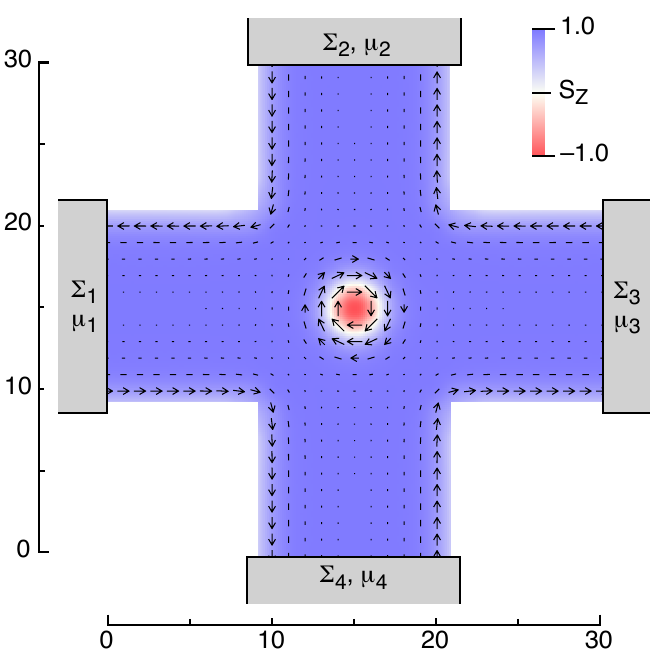}
\par\end{centering}

\protect\caption{(color online) The geometry of a $31\times31$ tight-binding cross
bar. The arrows denote the in-plane component of the magnetization
texture of a single skyrmion. The color plot demonstrates the $S_{z}$
component. The four terminals are numbered clock-wise.\label{fig:SpinTexture}}
\end{figure}

Due to the lack of periodicity, we apply the non-equilibrium Green's function
method (NEGF) to simulate the coherent transport of itinerant spins
traversing a single magnetic skyrmion \cite{datta_quantum_2005}.
The tight-binding
electron Hamiltonian we employ is,
\begin{equation}
\mathbf{H}_{\textrm{e}}=-J_{\textrm{H}}\sum_{i}c_{i}^{\dagger}\mathbf{\boldsymbol{\sigma}}_{i}c_{i}\cdot\mathbf{S}_{i}-t\sum_{\left\langle i,j\right\rangle }\left(c_{i}^{\dagger}c_{j}+\textrm{h.c.}\right),\label{eq:ElectronHamiltonian}
\end{equation}
where $\boldsymbol{\sigma}_{i}$ is the spin of itinerant electrons,
$J_{\textrm{H}}$ is the Hunds' rule coupling, $t$ is the nearest
neighbor hopping, and $\mathbf{S}_{i}$ is the local magnetization.
It has been previously discussed that the external magnetic field
does not contribute much to the Hall effect, therefore we neglect its
effect on the electron by taking the hopping parameter to be real \cite{nagaosa_anomalous_2010}.
%
Thus, the Hall signal observed in the following
calculations is purely from the emergent gauge field of the skyrmion.
The spin
texture $\left\{ \mathbf{S}_{i}\right\} $ contains a single skyrmion
located at the center of a 4-terminal cross bar (as shown in Fig.
\ref{fig:SpinTexture}).
This texture is fully damped using the Landau-Lifshitz-Gilbert (LLG) equation
with the magnetic Hamiltonian $H_{\textrm{S}}=-J\sum_{\left\langle i,j\right\rangle }\mathbf{S}_{i}\cdot\mathbf{S}_{j}-D\sum_{\left\langle i,j\right\rangle }\mathbf{\hat{\mathbf{r}}}_{i,j}\cdot\mathbf{S}_{i}\times\mathbf{S}_{j}-\sum_{i}\mathbf{h}_{0}\cdot\mathbf{S}_{i}$.
Here $J$ is the nearest neighbor exchange coupling, $D$ is the DM
interaction and $\mathbf{h}_{0}$ is the external magnetic field perpendicular
to the cross-bar plane. For simplicity we choose $D=J=h_{0}^{z}$.
Periodic magnetic boundaries are applied at the terminals, while open
magnetic boundaries are used for other boundaries along the cross-bar,
which gives large in-plane magnetization components at the edges.
The skyrmion at the center is generated by manually creating a unity
topological charge and then relaxing the spin texture until the magnetic
energy is stable.
%
Details of the magnetic dynamical simulations can be found in
Ref.\cite{yin_topological_2014}.

For the electron transport calculation, semi-infinite boundary
conditions for electron states are applied to the four terminals of
the cross bar.
Each terminal is assumed to be a thermal bath of carriers
with chemical potential $\mu_{m}$.
The semi-infinite electrodes are
included by adding self-energy terms, $\boldsymbol{\Sigma}_{m}=\mathbf{t}^{\dagger}\mathbf{g}_{m}^{R}\mathbf{t}$,
to the terminal blocks of $\mathbf{H}_{\textrm{e}}$, where $\mathbf{g}_{m}^{R}$
is the surface Green's function of terminal $m$.
The retarded Green's function
of the device region bounded by the terminals
is given by $\mathbf{G}^{R}=\left[\epsilon\mathbf{I}-\mathbf{H}_{\textrm{e}}-\sum_{m}\boldsymbol{\Sigma}_{m}\right]^{-1}$.
In the linear response limit, the zero-temperature terminal currents, $I_m$,
are given by
$I_m = (e/h) \sum_n T_{m,n} \delta \mu_n$.
$\delta \mu_n$
denotes the chemical potential shift due to an applied bias
in terminal $n$, ($\delta\mu_{n}=\mu_{n}-\epsilon_{F}$).
$T_{m,n}=\textrm{Tr}\left[\boldsymbol{\Gamma}_{m}\boldsymbol{G}_{mn}^{R}\boldsymbol{\Gamma}_{n}\boldsymbol{G}_{mn}^{A}\right]$
($m\neq n$) is the transmission coefficient between terminal $m$
and $n$, where $\mathbf{G}_{mn}^{A}=\mathbf{G}_{mn}^{R\dagger}$,
and
$\boldsymbol{\Gamma}_{m}=i\left(\boldsymbol{\Sigma}_{m}-\boldsymbol{\Sigma}_{m}^{\dagger}\right)$.
At steady state, the charge current is conserved
such that $T_{mm}=-\sum_{n\neq m}T_{mn}$.
A Symmetric bias is applied
between terminals $1$ and $3$, $\delta\mu_{1}=-\delta\mu_{3}=\delta\mu=0.1J_{\textrm{H}}$.
Enforcing $I_{2}=I_{4}=0$ in the Hall effect measurement, the transverse
Hall voltage can be solved as $\delta\mu_{2}=\delta\mu\left(P-Q\right)/\left(P+Q\right)$
and $\delta\mu_{4}=\delta\mu\left(R-S\right)/\left(R+S\right)$, where
\begin{equation}
\begin{cases}
\begin{array}{c}
P=T_{21}T_{41}+T_{21}T_{42}+T_{21}T_{43}+T_{24}T_{41}\\
Q=T_{23}T_{41}+T_{23}T_{42}+T_{23}T_{43}+T_{24}T_{43}\\
R=T_{42}T_{21}+T_{21}T_{41}+T_{23}T_{41}+T_{24}T_{41}\\
S=T_{42}T_{23}+T_{21}T_{43}+T_{23}T_{43}+T_{24}T_{43}
\end{array} & .\end{cases}\label{eq:HallSignalFactors}
\end{equation}
Thus, the topological Hall angle can be evaluated as
$\tan\theta_{\textrm{TH}} = E_H/E_x=(\mu_2-\mu_4)/(\mu_1-\mu_3)$.

Once $\delta\mu_m$ and $I_m$ are obtained, then
the total terminal spin current,
$I_{m}^{S_{\alpha}}$($\alpha=x,y,z$),
is evaluated from
$I_{m}^{S_{\alpha}}=\frac{\hbar}{2} \textrm{Tr}
\left[\boldsymbol{\sigma}_{\alpha}\mathbf{I}_{m}^{\textrm{neq}}\right],$
where
$\boldsymbol{\sigma}_{\alpha}=\mathcal{I}\otimes\sigma_{\alpha}$
is the extended Pauli matrix and
$\mathbf{I}_{m}^{\textrm{neq}}$
is the terminal current operator
$\mathbf{I}_{m}^{\textrm{neq}}=\frac{i}{2\pi\hbar}\left[\delta\mathbf{G}_{m}^{n}\boldsymbol{\Sigma}_{m}^{\dagger}-\boldsymbol{\Sigma}_{m}\delta\mathbf{G}_{m}^{n}\right.+\left.\mathbf{G}_{m}^{R}\delta\boldsymbol{\Sigma}_{m}^{\textrm{in}}-\delta\boldsymbol{\Sigma}_{m}^{\textrm{in}}\mathbf{G}_{m}^{A}\right],$
$\delta\mathbf{G}_{m}^{n} = \sum_n \mathbf{G}^R_{m,n}\boldsymbol{\Gamma}_{n,n} \mathbf{G}^A_{n,m} \delta \mu_n$, and
$\delta\boldsymbol{\Sigma}_{m}^{\textrm{in}} = \boldsymbol{\Gamma}_{m}\left(\epsilon_{F}\right)
\delta\mu_{m}$.
The intensity of the TSHE is described by
the spin Hall angle, a renormalized ratio between the transverse spin
current and the longitudinal charged current
\begin{equation}
\theta_{\textrm{TSH}}=\left(\frac{2e}{\hbar}\right)\frac{\sigma_{xy}^{S_{z}}}{\sigma_{xx}}=\left(\frac{2e}{\hbar}\right)\frac{I_{42}^{s_{z}}}{I_{13}},\label{eq:TSH_angle}
\end{equation}
where $I_{13}=I_{1}-I_{3}$, and $I_{42}^{S_{z}}=I_{4}^{S_{z}}-I_{2}^{S_{z}}$.

\begin{figure}
\begin{centering}
\includegraphics[width=0.83\columnwidth]{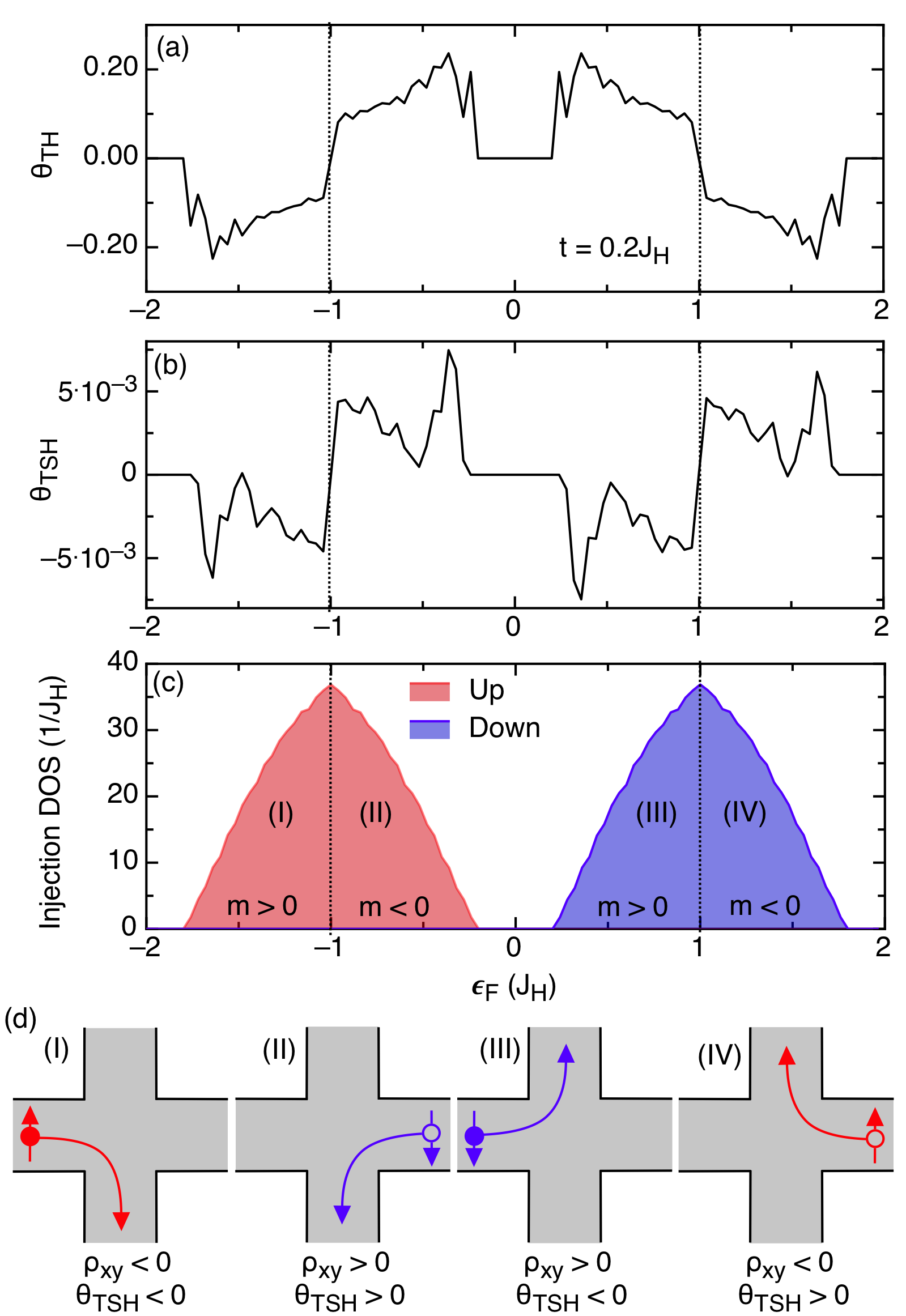}
\par\end{centering}
\protect\caption{
(color online) THE and TSHE for the case of pure spin injection ($t=0.2J_{\textrm{H}}$).
The (a) Hall angle $\theta_{\rm TH}$ and the (b) spin Hall
angle $\theta_{\rm TSH}$ are shown as a function of $\epsilon_{F}$.
The surface
density of states at terminal $1$ is shown in (c).
The four scenarios of
different carrier-type and spin compositions are illustrated in
(d).
}
\label{fig:PureSpinInjection}
\end{figure}

First, we study the THE and TSHE for the case of pure spin injection.
By setting $t=0.2J_{\textrm{H}}$,
the tight-binding band-width is smaller than the spin splitting given
by $J_{\textrm{H}}$.
Therefore no matter where the the Fermi level
lies, the electron injection does not mix different spins.
The $\theta_{\textrm{TH}}$ and $\theta_{\textrm{TSH}}$ for different positions of $\epsilon_{F}$
are shown in Fig. \ref{fig:PureSpinInjection}(a) and (b),
respectively.
The corresponding surface density of states (DOS) that
determines the type of current injection at terminal $1$ is shown
in Fig. \ref{fig:PureSpinInjection}(c).
When the surface DOS is zero, the Fermi surface lies in the spin
gap, and injection is absent, so that both $\theta_{\textrm{TH}}$ and $\theta_{\textrm{TSH}}$
are suppressed to zero.
As $\epsilon_{F}$ passes through the bands,
pure spin injection gives a Hall angle up to $\pm0.2$ indicating
the expected THE.
The corresponding value of $\theta_{\textrm{TSH}}$
is within $\pm0.005$.
At $\epsilon_{F}=\pm J_{\textrm{H}}$, both $\theta_{\textrm{TH}}$ and $\theta_{\textrm{TSH}}$
change sign.

The sign change of the Hall angles can be explained by the spin and carrier-type
composition of the injection from the ferromagnetic contacts.
For each transport channel, a one-dimensional tight-binding chain
gives a negative cosine electron band dispersion, which has a sign
change of the effective mass at the band center.
The effective mass ($m^{*}$) is positive at the bottom band-edge, and becomes negative at the top.
When an up-spin electron with positive
$m^{*}$ is injected from terminal $1$,
it is scattered to the ``right'' due to the effect of the emergent gauge field
generated by the skyrmion.
This is denoted as scenario (I) in Fig. \ref{fig:PureSpinInjection}(d).
Alternately when $m^{*}<0$, an up-spin electron injected from terminal
$1$ is equivalent to a down-spin hole injected from terminal $3$.
Since the spin scattering due to the skyrmion is anti-symmetric, the down-spin hole is scattered to its
``left'' as denoted by scenario (II).
In a multi-channel scenario due to the transverse confinement, the tight-binding band splits into several
sub-bands.
Thus, the number of the electron bands
and the hole bands crossing the Fermi level changes at different positions of $\epsilon_F$.
As $\epsilon_{F}$ moves
from the bottom band-edge to the band-center, the number of electron
bands crossing $\epsilon_{F}$ decreases, while the number of hole
bands increases as depicted in Fig. \ref{fig:PureSpinInjection}(c).
Right at the band-center, the electrons and holes are equal,
indicating an equal contribution from both scenarios (I) and (II),
which leads to a cancellation of both $\theta_{\textrm{TH}}$ and
$\theta_{\textrm{TSH}}$.
Further increasing $\epsilon_{F}$, scenario
(II) starts to dominate such that $\theta_{\textrm{TH}}$ and $\theta_{\textrm{TSH}}$
change sign. Similar arguments can be applied
to scenario (III) and (IV) for the down-spin case (see Fig. \ref{fig:PureSpinInjection}(d)).

Semiclassically, the relative strength of THE to the TSHE can understood as a
cancellation of the transverse electric field due to charge accumulation at contacts (2) and (4)
with the gauge field of the skyrmion.
In all these pure-spin injection scenarios, the spin current is carried
by charge which leads to a transverse accumulation of charge resulting in a
Hall voltage and hence a THE.
Since the transverse electric field cancels the Lorentz force given
by the emergent gauge field of the skyrmion, a continuous spin
current is suppressed at the steady state, making the TSHE
insignificant. However, an order-of-magnitude increase in
$\theta_{\textrm{TSH}}$ can be achieved in the case of mixed spin
injection which we discuss next.

\begin{figure}
\begin{centering}
\includegraphics[width=0.8\columnwidth]{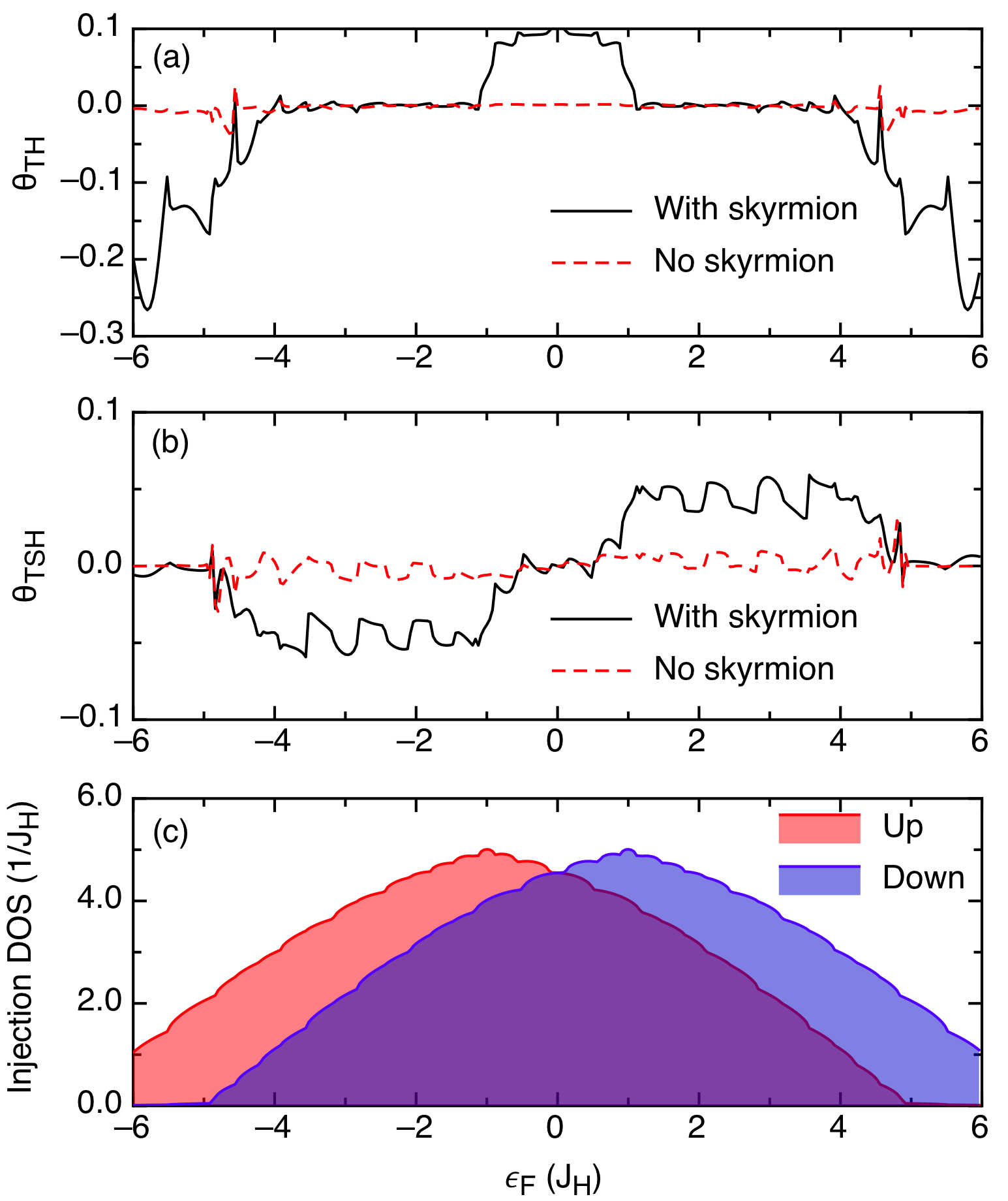}
\par\end{centering}
\protect\caption{(color online) THE and TSHE in the case of mixed spin injection ($t=1.5J_{\textrm{H}}$).
(a) and (b) demonstrate the values of  $\theta_{\textrm{TH}}$ and $\theta_{\textrm{TSH}}$
for different positions of $\epsilon_{F}$. The red dashed lines correspond
to the case where the central skyrmion is removed. (c) is a plot of
the surface DOS at terminal $1$.\label{fig:MixedSpinInjection}}
\end{figure}

To simulate mixed spin injection,
the hopping term is increased to $t=1.5J_{\textrm{H}}$ such that
the injection band-widths of each spin are enlarged and overlap
in some range of $\epsilon_{F}$.
The calculated values of the $\theta_{\textrm{TH}}$ and
$\theta_{\textrm{TSH}}$ are shown in Fig. \ref{fig:MixedSpinInjection},
along with the corresponding results in the absence of a skyrmion for comparison.
For energies in the range of $-4.5J_{\textrm{H}}<\epsilon_{F}<-J_{\textrm{H}}$
and $J_{\textrm{H}}<\epsilon_{F}<4.5J_{\textrm{H}}$, $\theta_{\textrm{TH}}$
vanishes to $\sim0$, whereas $\theta_{\textrm{TSH}}$ increases by
approximately an order of magnitude compared to the case of pure-spin injection.
Additionally, in the energy range $-J_{\textrm{H}} < \epsilon_{F} < J_{\textrm{H}}$, the
Hall angle corresponding to the THE $\theta_{\textrm{H}}$ is finite and roughly same order
as that in the case of pure-spin injection.

To explain the presence of the TSHE, we again refer to the
four scenarios shown in Fig. \ref{fig:PureSpinInjection}(d).
Within $-4.5J_{\textrm{H}}<\epsilon_{F}<-J_{\textrm{H}}$, the transport
is dominated by scenario (I)+(III) as shown in Fig. \ref{fig:MixedSpinInjection}(c).
In this case, both the spin-up and spin-down electrons are injected
from terminal $1$.
Due to the presence of a skyrmion there exists a topological Hall effect which
produces a transverse electrical field, $E_\textrm{TH}$.
At steady state,
the zero-current condition at terminals 2 and 4 requires $eE_\textrm{TH}=-F\uparrow$ and
$eE_\textrm{TH}=-F\downarrow$ satisfied simultaneously.
Due to the chirality of the skyrmion, the emergent field experienced by the up spin is
opposite to that experienced by the down spin, which generates opposite emergent
Lorentz forces on the two types of spins ($F\uparrow=-F\downarrow$).
Therefore, the zero-current condition in the transverse direction
cannot be satisfied unless $E_{\textrm{TH}}=0$.
Although imbalanced spin injection occurs
due to the ferromagnetic electrodes, the THE must be suppressed in steady state
as long as the transport is dominated by the same type of carrier.
Since there is no electrostatic field to balance the emergent Lorentz force,
a continuous chargeless spin current is established.
Similar
explanations {[}(II)+(IV){]} can be applied for $J_{\textrm{H}}<\epsilon_{F}<4.5J_{\textrm{H}}$.

When the transport is dominated by two different types of carriers
with the same spin,
the TSHE is suppressed, and the THE voltage becomes finite.
In our calculations, this occurs when $\epsilon_{F}$ is within $\left[-J_{\textrm{H}},J_{\textrm{H}}\right]$,
and the transport is dominated by the scenarios (II)+(III). In this case the down-spin electrons and
holes are injected from terminals $1$ and $3$, respectively.
The
electrons and holes are scattered in opposite directions and then accumulate
at terminals $2$ and $4$, respectively.
Since the same spin is assigned to
opposite charges, a non-zero $E_{\textrm{TH}}$ develops at terminals (2) and (4) resulting in a
finite THE with a vanishing TSHE.

\begin{figure}[h]
\begin{centering}
\includegraphics[width=0.8\columnwidth]{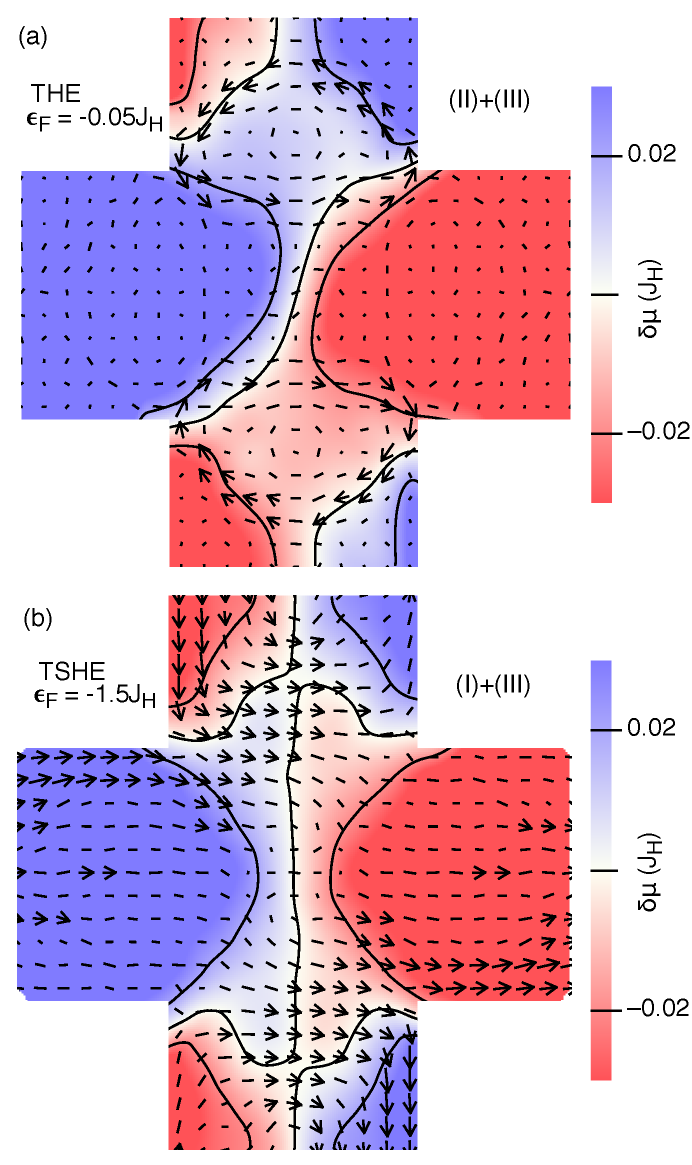}
\par\end{centering}
\protect\caption{
(color online)
Vector map of $\vec{J}_{S_{z}}$ (arrow plot) and the
effective chemical potential distribution (color map) for (a) the THE
and (b) the TSHE.
A longitudinal applied bias of $\delta\mu_{1}=-\delta\mu_{3}=0.1J_{\textrm{H}}$ is
applied.
For the THE (a), the spin current symmetrically circulates on
either side of the the skyrmion resulting in no
net transverse spin current.
The electron and hole accumulation induces an imbalanced transverse
potential distribution.
For the TSHE (b), the transverse
chemical potential distribution is symmetric,
and a charge-less spin current is established in the
transverse direction.
\label{fig:CurrentPotentialMap}}
\end{figure}

To further demonstrate the differences between the THE and the TSHE,
we show the vector map of the spin current density
$\mathbf{J}_{S_{z}}\left(\mathbf{r}\right)$ and the corresponding
color map of the charge accumulation in Fig.
\ref{fig:CurrentPotentialMap}.
The spin texture and the terminal numbering are the same as in Fig. \ref{fig:SpinTexture}.
For the THE case shown in Fig. \ref{fig:CurrentPotentialMap}(a),
$\epsilon_{F}=-0.05J_{\textrm{H}}$ and the transport
is dominated by scenario (II)+(III).
There is a net drop in the transverse chemical potential between leads 2 and 4.
%
The $\mathbf{J}_{S_{z}}$ vectors circulate symmetrically on either
side of the skyrmion, generating no significant total transverse
spin current.
This corresponds to the case where $\theta_{\textrm{TH}}\approx-0.2$
and $\theta_{\textrm{TSH}}\approx0$.
For the TSHE case shown in {[}Fig.
\ref{fig:CurrentPotentialMap}(b){]},
the transport is dominated by scenario (I)+(III).
The equal-potential contour of $\delta\mu\left(\mathbf{r}\right)=0$
cuts all the way across the vertical bar indicating little charge
imbalance between leads 2 and 4.
In transverse leads 2 and 4 there is a net spin current
directed from lead 2 to lead 4
giving a negative $\theta_{\textrm{TSH}}\approx-0.05$.

%
%
The TSHE discussed here is of similar magnitude as the SHE in broadly used Pt thin films \cite{liu_spin-torque_2011}. 
However, the physical mechanism giving rise to the TSHE is fundamentally different from the one leading to the spin
Hall effect in strong spin orbit coupled systems. 
In such systems, the spin Hall effect results from the topological property of the Bloch bands in momentum space. 
In contrast the TSHE results from the topological property of the skyrmion spin texture in real space. 
The real-space topology exerts opposite emergent Lorentz forces on different spins, which can induce the TSHE. 
%
%
%

We thank the helpful discussions with Prof. Jing Shi at UCR and Dr.
K. M. Masum Habib at Univ. of Virginia. 
This work was supported by the NSF (ECCS-1408168).

\bibliographystyle{apsrev}

\begin{thebibliography}{30}
\expandafter\ifx\csname natexlab\endcsname\relax\def\natexlab#1{#1}\fi
\expandafter\ifx\csname bibnamefont\endcsname\relax
  \def\bibnamefont#1{#1}\fi
\expandafter\ifx\csname bibfnamefont\endcsname\relax
  \def\bibfnamefont#1{#1}\fi
\expandafter\ifx\csname citenamefont\endcsname\relax
  \def\citenamefont#1{#1}\fi
\expandafter\ifx\csname url\endcsname\relax
  \def\url#1{\texttt{#1}}\fi
\expandafter\ifx\csname urlprefix\endcsname\relax\def\urlprefix{URL }\fi
\providecommand{\bibinfo}[2]{#2}
\providecommand{\eprint}[2][]{\url{#2}}

\bibitem[{\citenamefont{Dyakonov and
  Perel}(1971{\natexlab{a}})}]{dyakonov_current-induced_1971}
\bibinfo{author}{\bibfnamefont{M.~I.} \bibnamefont{Dyakonov}} \bibnamefont{and}
  \bibinfo{author}{\bibfnamefont{V.~I.} \bibnamefont{Perel}},
  \bibinfo{journal}{Physics Letters A} \textbf{\bibinfo{volume}{35}},
  \bibinfo{pages}{459} (\bibinfo{year}{1971}{\natexlab{a}}), ISSN
  \bibinfo{issn}{0375-9601},
  \urlprefix\url{http://www.sciencedirect.com/science/article/pii/0375960171901964}.

\bibitem[{\citenamefont{Dyakonov and
  Perel}(1971{\natexlab{b}})}]{dyakonov_possibility_1971}
\bibinfo{author}{\bibfnamefont{M.~I.} \bibnamefont{Dyakonov}} \bibnamefont{and}
  \bibinfo{author}{\bibfnamefont{V.~I.} \bibnamefont{Perel}},
  \bibinfo{journal}{Sov. Phys. JETP Lett.} \textbf{\bibinfo{volume}{13}},
  \bibinfo{pages}{467} (\bibinfo{year}{1971}{\natexlab{b}}),
  \urlprefix\url{http://www.jetpletters.ac.ru/ps/1587/article_24366.shtml}.

\bibitem[{\citenamefont{Bakun et~al.}(1984)\citenamefont{Bakun, Zakharchenya,
  Rogachev, Tkachuk, and Fleisher}}]{bakun_observation_1984}
\bibinfo{author}{\bibfnamefont{A.~A.} \bibnamefont{Bakun}},
  \bibinfo{author}{\bibfnamefont{B.~P.} \bibnamefont{Zakharchenya}},
  \bibinfo{author}{\bibfnamefont{A.~A.} \bibnamefont{Rogachev}},
  \bibinfo{author}{\bibfnamefont{M.~N.} \bibnamefont{Tkachuk}},
  \bibnamefont{and} \bibinfo{author}{\bibfnamefont{V.~G.}
  \bibnamefont{Fleisher}}, \bibinfo{journal}{Sov. Phys. JETP Lett.}
  (\bibinfo{year}{1984}).

\bibitem[{\citenamefont{Tkachuk et~al.}(1986)\citenamefont{Tkachuk,
  Zakharchenya, and Fleisher}}]{tkachuk_resonant_1986}
\bibinfo{author}{\bibfnamefont{M.~N.} \bibnamefont{Tkachuk}},
  \bibinfo{author}{\bibfnamefont{B.~P.} \bibnamefont{Zakharchenya}},
  \bibnamefont{and} \bibinfo{author}{\bibfnamefont{V.~G.}
  \bibnamefont{Fleisher}}, \bibinfo{journal}{Sov. Phys. JETP Lett.}
  (\bibinfo{year}{1986}).

\bibitem[{\citenamefont{Hirsch}(1999)}]{hirsch_spin_1999}
\bibinfo{author}{\bibfnamefont{J.~E.} \bibnamefont{Hirsch}},
  \bibinfo{journal}{Phys. Rev. Lett.} \textbf{\bibinfo{volume}{83}},
  \bibinfo{pages}{1834} (\bibinfo{year}{1999}),
  \urlprefix\url{http://link.aps.org/doi/10.1103/PhysRevLett.83.1834}.

\bibitem[{\citenamefont{Murakami et~al.}(2003)\citenamefont{Murakami, Nagaosa,
  and Zhang}}]{murakami_dissipationless_2003}
\bibinfo{author}{\bibfnamefont{S.}~\bibnamefont{Murakami}},
  \bibinfo{author}{\bibfnamefont{N.}~\bibnamefont{Nagaosa}}, \bibnamefont{and}
  \bibinfo{author}{\bibfnamefont{S.-C.} \bibnamefont{Zhang}},
  \bibinfo{journal}{Science} \textbf{\bibinfo{volume}{301}},
  \bibinfo{pages}{1348} (\bibinfo{year}{2003}), ISSN \bibinfo{issn}{0036-8075,
  1095-9203}, \urlprefix\url{http://www.sciencemag.org/content/301/5638/1348}.

\bibitem[{\citenamefont{Sinova et~al.}(2004)\citenamefont{Sinova, Culcer, Niu,
  Sinitsyn, Jungwirth, and MacDonald}}]{sinova_universal_2004}
\bibinfo{author}{\bibfnamefont{J.}~\bibnamefont{Sinova}},
  \bibinfo{author}{\bibfnamefont{D.}~\bibnamefont{Culcer}},
  \bibinfo{author}{\bibfnamefont{Q.}~\bibnamefont{Niu}},
  \bibinfo{author}{\bibfnamefont{N.~A.} \bibnamefont{Sinitsyn}},
  \bibinfo{author}{\bibfnamefont{T.}~\bibnamefont{Jungwirth}},
  \bibnamefont{and} \bibinfo{author}{\bibfnamefont{A.~H.}
  \bibnamefont{MacDonald}}, \bibinfo{journal}{Phys. Rev. Lett.}
  \textbf{\bibinfo{volume}{92}}, \bibinfo{pages}{126603}
  (\bibinfo{year}{2004}),
  \urlprefix\url{http://link.aps.org/doi/10.1103/PhysRevLett.92.126603}.

\bibitem[{\citenamefont{Nikoli{\'c} et~al.}(2005)\citenamefont{Nikoli{\'c},
  Souma, Z{\^a}rbo, and Sinova}}]{nikolic_nonequilibrium_2005}
\bibinfo{author}{\bibfnamefont{B.~K.} \bibnamefont{Nikoli{\'c}}},
  \bibinfo{author}{\bibfnamefont{S.}~\bibnamefont{Souma}},
  \bibinfo{author}{\bibfnamefont{L.~P.} \bibnamefont{Z{\^a}rbo}},
  \bibnamefont{and} \bibinfo{author}{\bibfnamefont{J.}~\bibnamefont{Sinova}},
  \bibinfo{journal}{Phys. Rev. Lett.} \textbf{\bibinfo{volume}{95}},
  \bibinfo{pages}{046601} (\bibinfo{year}{2005}),
  \urlprefix\url{http://link.aps.org/doi/10.1103/PhysRevLett.95.046601}.

\bibitem[{\citenamefont{Grover and Senthil}(2008)}]{grover_topological_2008}
\bibinfo{author}{\bibfnamefont{T.}~\bibnamefont{Grover}} \bibnamefont{and}
  \bibinfo{author}{\bibfnamefont{T.}~\bibnamefont{Senthil}},
  \bibinfo{journal}{Phys. Rev. Lett.} \textbf{\bibinfo{volume}{100}},
  \bibinfo{pages}{156804} (\bibinfo{year}{2008}),
  \urlprefix\url{http://link.aps.org/doi/10.1103/PhysRevLett.100.156804}.

\bibitem[{\citenamefont{Iwasaki et~al.}(2013)\citenamefont{Iwasaki, Mochizuki,
  and Nagaosa}}]{iwasaki_universal_2013}
\bibinfo{author}{\bibfnamefont{J.}~\bibnamefont{Iwasaki}},
  \bibinfo{author}{\bibfnamefont{M.}~\bibnamefont{Mochizuki}},
  \bibnamefont{and} \bibinfo{author}{\bibfnamefont{N.}~\bibnamefont{Nagaosa}},
  \bibinfo{journal}{Nature Communications} \textbf{\bibinfo{volume}{4}},
  \bibinfo{pages}{1463} (\bibinfo{year}{2013}), ISSN \bibinfo{issn}{2041-1723},
  \urlprefix\url{http://www.nature.com/ncomms/journal/v4/n2/ncomms2442/metrics}.

\bibitem[{\citenamefont{Wunderlich et~al.}(2005)\citenamefont{Wunderlich,
  Kaestner, Sinova, and Jungwirth}}]{wunderlich_experimental_2005}
\bibinfo{author}{\bibfnamefont{J.}~\bibnamefont{Wunderlich}},
  \bibinfo{author}{\bibfnamefont{B.}~\bibnamefont{Kaestner}},
  \bibinfo{author}{\bibfnamefont{J.}~\bibnamefont{Sinova}}, \bibnamefont{and}
  \bibinfo{author}{\bibfnamefont{T.}~\bibnamefont{Jungwirth}},
  \bibinfo{journal}{Phys. Rev. Lett.} \textbf{\bibinfo{volume}{94}},
  \bibinfo{pages}{047204} (\bibinfo{year}{2005}),
  \urlprefix\url{http://link.aps.org/doi/10.1103/PhysRevLett.94.047204}.

\bibitem[{\citenamefont{Kato et~al.}(2004)\citenamefont{Kato, Myers, Gossard,
  and Awschalom}}]{kato_observation_2004}
\bibinfo{author}{\bibfnamefont{Y.~K.} \bibnamefont{Kato}},
  \bibinfo{author}{\bibfnamefont{R.~C.} \bibnamefont{Myers}},
  \bibinfo{author}{\bibfnamefont{A.~C.} \bibnamefont{Gossard}},
  \bibnamefont{and} \bibinfo{author}{\bibfnamefont{D.~D.}
  \bibnamefont{Awschalom}}, \bibinfo{journal}{Science}
  \textbf{\bibinfo{volume}{306}}, \bibinfo{pages}{1910} (\bibinfo{year}{2004}),
  ISSN \bibinfo{issn}{0036-8075, 1095-9203},
  \urlprefix\url{http://www.sciencemag.org/content/306/5703/1910}.

\bibitem[{\citenamefont{Smit}(1955)}]{smit_spontaneous_1955}
\bibinfo{author}{\bibfnamefont{J.}~\bibnamefont{Smit}},
  \bibinfo{journal}{Physica} \textbf{\bibinfo{volume}{21}},
  \bibinfo{pages}{877} (\bibinfo{year}{1955}), ISSN \bibinfo{issn}{0031-8914},
  \urlprefix\url{http://www.sciencedirect.com/science/article/pii/S0031891455925969}.

\bibitem[{\citenamefont{Berger}(1970)}]{berger_side-jump_1970}
\bibinfo{author}{\bibfnamefont{L.}~\bibnamefont{Berger}},
  \bibinfo{journal}{Phys. Rev. B} \textbf{\bibinfo{volume}{2}},
  \bibinfo{pages}{4559} (\bibinfo{year}{1970}),
  \urlprefix\url{http://link.aps.org/doi/10.1103/PhysRevB.2.4559}.

\bibitem[{\citenamefont{Onoda and Nagaosa}(2002)}]{onoda_topological_2002}
\bibinfo{author}{\bibfnamefont{M.}~\bibnamefont{Onoda}} \bibnamefont{and}
  \bibinfo{author}{\bibfnamefont{N.}~\bibnamefont{Nagaosa}},
  \bibinfo{journal}{J. Phys. Soc. Jpn.} \textbf{\bibinfo{volume}{71}},
  \bibinfo{pages}{19} (\bibinfo{year}{2002}), ISSN \bibinfo{issn}{0031-9015},
  \urlprefix\url{http://journals.jps.jp/doi/abs/10.1143/JPSJ.71.19}.

\bibitem[{\citenamefont{Nagaosa et~al.}(2010)\citenamefont{Nagaosa, Sinova,
  Onoda, MacDonald, and Ong}}]{nagaosa_anomalous_2010}
\bibinfo{author}{\bibfnamefont{N.}~\bibnamefont{Nagaosa}},
  \bibinfo{author}{\bibfnamefont{J.}~\bibnamefont{Sinova}},
  \bibinfo{author}{\bibfnamefont{S.}~\bibnamefont{Onoda}},
  \bibinfo{author}{\bibfnamefont{A.~H.} \bibnamefont{MacDonald}},
  \bibnamefont{and} \bibinfo{author}{\bibfnamefont{N.~P.} \bibnamefont{Ong}},
  \bibinfo{journal}{Rev. Mod. Phys.} \textbf{\bibinfo{volume}{82}},
  \bibinfo{pages}{1539} (\bibinfo{year}{2010}),
  \urlprefix\url{http://link.aps.org/doi/10.1103/RevModPhys.82.1539}.

\bibitem[{\citenamefont{Haldane}(2004)}]{haldane_berry_2004}
\bibinfo{author}{\bibfnamefont{F.~D.~M.} \bibnamefont{Haldane}},
  \bibinfo{journal}{Phys. Rev. Lett.} \textbf{\bibinfo{volume}{93}},
  \bibinfo{pages}{206602} (\bibinfo{year}{2004}),
  \urlprefix\url{http://link.aps.org/doi/10.1103/PhysRevLett.93.206602}.

\bibitem[{\citenamefont{Volovik}(1989)}]{volovik_fractional_1989}
\bibinfo{author}{\bibfnamefont{G.~E.} \bibnamefont{Volovik}}, in
  \emph{\bibinfo{booktitle}{{AIP} {Conference} {Proceedings}}}
  (\bibinfo{publisher}{AIP Publishing}, \bibinfo{year}{1989}), vol.
  \bibinfo{volume}{194}, pp. \bibinfo{pages}{136--146},
  \urlprefix\url{http://scitation.aip.org/content/aip/proceeding/aipcp/10.1063/1.38806}.

\bibitem[{\citenamefont{Taguchi et~al.}(2001)\citenamefont{Taguchi, Oohara,
  Yoshizawa, Nagaosa, and Tokura}}]{taguchi_spin_2001}
\bibinfo{author}{\bibfnamefont{Y.}~\bibnamefont{Taguchi}},
  \bibinfo{author}{\bibfnamefont{Y.}~\bibnamefont{Oohara}},
  \bibinfo{author}{\bibfnamefont{H.}~\bibnamefont{Yoshizawa}},
  \bibinfo{author}{\bibfnamefont{N.}~\bibnamefont{Nagaosa}}, \bibnamefont{and}
  \bibinfo{author}{\bibfnamefont{Y.}~\bibnamefont{Tokura}},
  \bibinfo{journal}{Science} \textbf{\bibinfo{volume}{291}},
  \bibinfo{pages}{2573} (\bibinfo{year}{2001}), ISSN \bibinfo{issn}{0036-8075,
  1095-9203}, \urlprefix\url{http://www.sciencemag.org/content/291/5513/2573}.

\bibitem[{\citenamefont{Bruno et~al.}(2004)\citenamefont{Bruno, Dugaev, and
  Taillefumier}}]{bruno_topological_2004}
\bibinfo{author}{\bibfnamefont{P.}~\bibnamefont{Bruno}},
  \bibinfo{author}{\bibfnamefont{V.~K.} \bibnamefont{Dugaev}},
  \bibnamefont{and}
  \bibinfo{author}{\bibfnamefont{M.}~\bibnamefont{Taillefumier}},
  \bibinfo{journal}{Phys. Rev. Lett.} \textbf{\bibinfo{volume}{93}},
  \bibinfo{pages}{096806} (\bibinfo{year}{2004}),
  \urlprefix\url{http://link.aps.org/doi/10.1103/PhysRevLett.93.096806}.

\bibitem[{\citenamefont{Yu et~al.}(2010)\citenamefont{Yu, Onose, Kanazawa,
  Park, Han, Matsui, Nagaosa, and Tokura}}]{yu_real-space_2010}
\bibinfo{author}{\bibfnamefont{X.~Z.} \bibnamefont{Yu}},
  \bibinfo{author}{\bibfnamefont{Y.}~\bibnamefont{Onose}},
  \bibinfo{author}{\bibfnamefont{N.}~\bibnamefont{Kanazawa}},
  \bibinfo{author}{\bibfnamefont{J.~H.} \bibnamefont{Park}},
  \bibinfo{author}{\bibfnamefont{J.~H.} \bibnamefont{Han}},
  \bibinfo{author}{\bibfnamefont{Y.}~\bibnamefont{Matsui}},
  \bibinfo{author}{\bibfnamefont{N.}~\bibnamefont{Nagaosa}}, \bibnamefont{and}
  \bibinfo{author}{\bibfnamefont{Y.}~\bibnamefont{Tokura}},
  \bibinfo{journal}{Nature} \textbf{\bibinfo{volume}{465}},
  \bibinfo{pages}{901} (\bibinfo{year}{2010}), ISSN \bibinfo{issn}{0028-0836},
  \urlprefix\url{http://www.nature.com/nature/journal/v465/n7300/full/nature09124.html}.

\bibitem[{\citenamefont{M{\"u}hlbauer et~al.}(2009)\citenamefont{M{\"u}hlbauer,
  Binz, Jonietz, Pfleiderer, Rosch, Neubauer, Georgii, and
  B{\"o}ni}}]{muhlbauer_skyrmion_2009}
\bibinfo{author}{\bibfnamefont{S.}~\bibnamefont{M{\"u}hlbauer}},
  \bibinfo{author}{\bibfnamefont{B.}~\bibnamefont{Binz}},
  \bibinfo{author}{\bibfnamefont{F.}~\bibnamefont{Jonietz}},
  \bibinfo{author}{\bibfnamefont{C.}~\bibnamefont{Pfleiderer}},
  \bibinfo{author}{\bibfnamefont{A.}~\bibnamefont{Rosch}},
  \bibinfo{author}{\bibfnamefont{A.}~\bibnamefont{Neubauer}},
  \bibinfo{author}{\bibfnamefont{R.}~\bibnamefont{Georgii}}, \bibnamefont{and}
  \bibinfo{author}{\bibfnamefont{P.}~\bibnamefont{B{\"o}ni}},
  \bibinfo{journal}{Science} \textbf{\bibinfo{volume}{323}},
  \bibinfo{pages}{915} (\bibinfo{year}{2009}), ISSN \bibinfo{issn}{0036-8075,
  1095-9203}, \urlprefix\url{http://www.sciencemag.org/content/323/5916/915}.

\bibitem[{\citenamefont{Schulz et~al.}(2012)\citenamefont{Schulz, Ritz, Bauer,
  Halder, Wagner, Franz, Pfleiderer, Everschor, Garst, and
  Rosch}}]{schulz_emergent_2012}
\bibinfo{author}{\bibfnamefont{T.}~\bibnamefont{Schulz}},
  \bibinfo{author}{\bibfnamefont{R.}~\bibnamefont{Ritz}},
  \bibinfo{author}{\bibfnamefont{A.}~\bibnamefont{Bauer}},
  \bibinfo{author}{\bibfnamefont{M.}~\bibnamefont{Halder}},
  \bibinfo{author}{\bibfnamefont{M.}~\bibnamefont{Wagner}},
  \bibinfo{author}{\bibfnamefont{C.}~\bibnamefont{Franz}},
  \bibinfo{author}{\bibfnamefont{C.}~\bibnamefont{Pfleiderer}},
  \bibinfo{author}{\bibfnamefont{K.}~\bibnamefont{Everschor}},
  \bibinfo{author}{\bibfnamefont{M.}~\bibnamefont{Garst}}, \bibnamefont{and}
  \bibinfo{author}{\bibfnamefont{A.}~\bibnamefont{Rosch}},
  \bibinfo{journal}{Nat Phys} \textbf{\bibinfo{volume}{8}},
  \bibinfo{pages}{301} (\bibinfo{year}{2012}), ISSN \bibinfo{issn}{1745-2473},
  \urlprefix\url{http://www.nature.com/nphys/journal/v8/n4/full/nphys2231.html?WT.ec_id=NPHYS-201204#ref5}.

\bibitem[{\citenamefont{Li et~al.}(2013)\citenamefont{Li, Kanazawa, Yu,
  Tsukazaki, Kawasaki, Ichikawa, Jin, Kagawa, and Tokura}}]{li_robust_2013}
\bibinfo{author}{\bibfnamefont{Y.}~\bibnamefont{Li}},
  \bibinfo{author}{\bibfnamefont{N.}~\bibnamefont{Kanazawa}},
  \bibinfo{author}{\bibfnamefont{X.~Z.} \bibnamefont{Yu}},
  \bibinfo{author}{\bibfnamefont{A.}~\bibnamefont{Tsukazaki}},
  \bibinfo{author}{\bibfnamefont{M.}~\bibnamefont{Kawasaki}},
  \bibinfo{author}{\bibfnamefont{M.}~\bibnamefont{Ichikawa}},
  \bibinfo{author}{\bibfnamefont{X.~F.} \bibnamefont{Jin}},
  \bibinfo{author}{\bibfnamefont{F.}~\bibnamefont{Kagawa}}, \bibnamefont{and}
  \bibinfo{author}{\bibfnamefont{Y.}~\bibnamefont{Tokura}},
  \bibinfo{journal}{Phys. Rev. Lett.} \textbf{\bibinfo{volume}{110}},
  \bibinfo{pages}{117202} (\bibinfo{year}{2013}),
  \urlprefix\url{http://link.aps.org/doi/10.1103/PhysRevLett.110.117202}.

\bibitem[{\citenamefont{Kanazawa et~al.}(2011)\citenamefont{Kanazawa, Onose,
  Arima, Okuyama, Ohoyama, Wakimoto, Kakurai, Ishiwata, and
  Tokura}}]{kanazawa_large_2011}
\bibinfo{author}{\bibfnamefont{N.}~\bibnamefont{Kanazawa}},
  \bibinfo{author}{\bibfnamefont{Y.}~\bibnamefont{Onose}},
  \bibinfo{author}{\bibfnamefont{T.}~\bibnamefont{Arima}},
  \bibinfo{author}{\bibfnamefont{D.}~\bibnamefont{Okuyama}},
  \bibinfo{author}{\bibfnamefont{K.}~\bibnamefont{Ohoyama}},
  \bibinfo{author}{\bibfnamefont{S.}~\bibnamefont{Wakimoto}},
  \bibinfo{author}{\bibfnamefont{K.}~\bibnamefont{Kakurai}},
  \bibinfo{author}{\bibfnamefont{S.}~\bibnamefont{Ishiwata}}, \bibnamefont{and}
  \bibinfo{author}{\bibfnamefont{Y.}~\bibnamefont{Tokura}},
  \bibinfo{journal}{Phys. Rev. Lett.} \textbf{\bibinfo{volume}{106}},
  \bibinfo{pages}{156603} (\bibinfo{year}{2011}),
  \urlprefix\url{http://link.aps.org/doi/10.1103/PhysRevLett.106.156603}.

\bibitem[{\citenamefont{Neubauer et~al.}(2009)\citenamefont{Neubauer,
  Pfleiderer, Binz, Rosch, Ritz, Niklowitz, and
  B{\"o}ni}}]{neubauer_topological_2009}
\bibinfo{author}{\bibfnamefont{A.}~\bibnamefont{Neubauer}},
  \bibinfo{author}{\bibfnamefont{C.}~\bibnamefont{Pfleiderer}},
  \bibinfo{author}{\bibfnamefont{B.}~\bibnamefont{Binz}},
  \bibinfo{author}{\bibfnamefont{A.}~\bibnamefont{Rosch}},
  \bibinfo{author}{\bibfnamefont{R.}~\bibnamefont{Ritz}},
  \bibinfo{author}{\bibfnamefont{P.~G.} \bibnamefont{Niklowitz}},
  \bibnamefont{and} \bibinfo{author}{\bibfnamefont{P.}~\bibnamefont{B{\"o}ni}},
  \bibinfo{journal}{Phys. Rev. Lett.} \textbf{\bibinfo{volume}{102}},
  \bibinfo{pages}{186602} (\bibinfo{year}{2009}),
  \urlprefix\url{http://link.aps.org/doi/10.1103/PhysRevLett.102.186602}.

\bibitem[{\citenamefont{Huang and Chien}(2012)}]{huang_extended_2012}
\bibinfo{author}{\bibfnamefont{S.~X.} \bibnamefont{Huang}} \bibnamefont{and}
  \bibinfo{author}{\bibfnamefont{C.~L.} \bibnamefont{Chien}},
  \bibinfo{journal}{Phys. Rev. Lett.} \textbf{\bibinfo{volume}{108}},
  \bibinfo{pages}{267201} (\bibinfo{year}{2012}),
  \urlprefix\url{http://link.aps.org/doi/10.1103/PhysRevLett.108.267201}.

\bibitem[{\citenamefont{Datta}(2005)}]{datta_quantum_2005}
\bibinfo{author}{\bibfnamefont{S.}~\bibnamefont{Datta}},
  \emph{\bibinfo{title}{Quantum {Transport}: {Atom} to {Transistor}}}
  (\bibinfo{publisher}{Cambridge University Press},
  \bibinfo{address}{Cambridge, UK; New York}, \bibinfo{year}{2005}),
  \bibinfo{edition}{2nd} ed., ISBN \bibinfo{isbn}{9780521631457}.

\bibitem[{\citenamefont{Yin et~al.}(2014)\citenamefont{Yin, Li, Kong, Lake,
  Chien, and Zang}}]{yin_topological_2014}
\bibinfo{author}{\bibfnamefont{G.}~\bibnamefont{Yin}},
  \bibinfo{author}{\bibfnamefont{Y.}~\bibnamefont{Li}},
  \bibinfo{author}{\bibfnamefont{L.}~\bibnamefont{Kong}},
  \bibinfo{author}{\bibfnamefont{R.~K.} \bibnamefont{Lake}},
  \bibinfo{author}{\bibfnamefont{C.~L.} \bibnamefont{Chien}}, \bibnamefont{and}
  \bibinfo{author}{\bibfnamefont{J.}~\bibnamefont{Zang}},
  \bibinfo{journal}{arXiv:1411.7762 [cond-mat]}  (\bibinfo{year}{2014}),
  \bibinfo{note}{arXiv: 1411.7762},
  \urlprefix\url{http://arxiv.org/abs/1411.7762}.

\bibitem[{\citenamefont{Liu et~al.}(2011)\citenamefont{Liu, Moriyama, Ralph,
  and Buhrman}}]{liu_spin-torque_2011}
\bibinfo{author}{\bibfnamefont{L.}~\bibnamefont{Liu}},
  \bibinfo{author}{\bibfnamefont{T.}~\bibnamefont{Moriyama}},
  \bibinfo{author}{\bibfnamefont{D.~C.} \bibnamefont{Ralph}}, \bibnamefont{and}
  \bibinfo{author}{\bibfnamefont{R.~A.} \bibnamefont{Buhrman}},
  \bibinfo{journal}{Phys. Rev. Lett.} \textbf{\bibinfo{volume}{106}},
  \bibinfo{pages}{036601} (\bibinfo{year}{2011}),
  \urlprefix\url{http://link.aps.org/doi/10.1103/PhysRevLett.106.036601}.

\end{thebibliography}

\end{document}